\documentclass[12pt]{article}
\usepackage{epsfig}

\usepackage{slashed}
\newcommand{\agt}{\,\rlap{\lower 3.5 pt \hbox{$\mathchar \sim$}} \raise 1pt
 \hbox {$>$}\,}
\newcommand{\alt}{\,\rlap{\lower 3.5 pt \hbox{$\mathchar \sim$}} \raise 1pt
 \hbox {$<$}\,}
\newcommand{\re}{\mathop{\mathrm{Re}}\nolimits}

\begin{document}

\title{
\vskip-3cm{\baselineskip14pt
\centerline{\normalsize DESY 08-010\hfill ISSN 0418-9833}
\centerline{\normalsize TTP08-05\hfill}
\centerline{\normalsize SFB/CPP-08-10\hfill}
\centerline{\normalsize January 2008\hfill}}
\vskip1.5cm
\bf Dominant two-loop electroweak corrections to the hadroproduction of a
pseudoscalar Higgs boson and its photonic decay}

\author{Joachim Brod$^*$, Frank Fugel$^\dagger$, and Bernd A.
Kniehl$^\ddagger$\\
\\
{\normalsize\it $^*$ Institut f\"ur Theoretische Teilchenphysik,
Universit\"at Karlsruhe,}\\
{\normalsize\it Engesserstra\ss e 7, 76131 Karlsruhe, Germany}\\
\\
{\normalsize\it $^\dagger$ Paul Scherrer Institut, 5232 Villigen PSI, Switzerland}\\
\\
{\normalsize\it $^\ddagger$ II. Institut f\"ur Theoretische Physik, Universit\"at Hamburg,}\\
{\normalsize\it Luruper Chaussee 149, 22761 Hamburg, Germany}}

\date{}

\maketitle

\begin{abstract}
We present the dominant two-loop electroweak corrections to the partial decay
widths to gluon jets and prompt photons of the neutral CP-odd Higgs boson
$A^0$, with mass $M_{A^0}<2M_W$, in the two-Higgs-doublet model for low to
intermediate values of the ratio $\tan\beta=v_2/v_1$ of the vacuum expectation
values.
They apply as they stand to the production cross sections in hadronic and
two-photon collisions, at the Tevatron, the LHC, and a future photon collider.
The appearance of three $\gamma_5$ matrices in closed fermion loops requires
special care in the dimensional regularization of ultraviolet divergences.
The corrections are negative and amount to several percent, so that they
fully compensate or partly screen the enhancement due to QCD corrections.

\medskip

\end{abstract}

\newpage

The search for Higgs bosons is among the prime tasks at the Fermilab Tevatron
and will be so at the CERN Large Hadron Collider (LHC), to go into operation
later during this year, and the International $e^+e^-$ Linear Collider (ILC),
which is currently being designed.
The standard model (SM) contains one complex Higgs doublet, from which
one neutral CP-even Higgs boson ($H$) emerges in the physical particle
spectrum after the electroweak symmetry breaking.
Despite its enormous success in describing almost all experimental particle
physics data available today, the SM is widely believed to be an effective
field theory, valid only at presently accessible energy scales, mainly because
of the naturalness problem related to the fine-tuning of the cut-off scale
appearing quadratically in the Higgs-boson mass counterterm, the failure of
gauge coupling unification, the absence of a concept to incorporate gravity,
and the lack of a cold-dark-matter candidate.
Supersymmetry (SUSY), which postulates the existence of a partner, with spin
shifted by half a unit, to each of the established matter and exchange
particles, is commonly viewed as the most attractive extension of the SM
solving all these problems.
The Higgs sector of the minimal SUSY extension of the SM (MSSM) consists of a
two-Higgs-doublet model (2HDM) and accommodates five physical Higgs bosons:
the neutral CP-even $h^0$ and $H^0$ bosons, the neutral CP-odd $A^0$ boson,
and the charged $H^\pm$-boson pair.
At the tree level, the MSSM Higgs sector has two free parameters, which are
usually taken to be the mass $M_{A^0}$ of the $A^0$ boson and the ratio
$\tan\beta=v_2/v_1$ of the vacuum expectation values of the two Higgs
doublets.

The discovery of the $A^0$ boson would rule out the SM and, at the same time,
give strong support to the MSSM.
At the LHC, this will be feasible except in the wedge of parameter space with
$M_{A^0}\agt250$~GeV and moderate value of $\tan\beta$, where only the $h^0$
boson can be detected \cite{Gianotti:2002xx}.
For low to intermediate values of $\tan\beta$, gluon fusion is by far the
dominant hadroproduction mechanism.
At large values of $\tan\beta$, $A^0b\overline{b}$ associated production
becomes important, too, especially at LHC c.m.\ energy, $\sqrt s=14$~TeV
\cite{Spira:1997dg}.
At the ILC operated in the $\gamma\gamma$ mode, via Compton back-scattering of
highly energetic laser light off the lepton beams, single production of the
$A^0$ boson will allow for its discovery, also throughout a large fraction of
the LHC wedge, and for a precision determination of its profile
\cite{Muhlleitner:2001kw}.
Two-photon collisions, albeit with less luminosity, will also take place in
the regular $e^+e^-$ mode of the ILC through electromagnetic bremsstrahlung or
beamstrahlung off the lepton beams.

In the mass range $M_{A^0}<2m_t$ and for large values of $\tan\beta$ in the
whole $M_{A^0}$ range, the $A^0$ boson dominantly decays to a $b\overline{b}$
pair, with a branching fraction of about 90\% \cite{Spira:1997dg}.
As in the case of the $H$ boson of the SM, the rare $\gamma\gamma$ decay
channel may then provide a useful signature at the LHC if the $b$ and
$\overline{b}$ quarks cannot be separated sufficiently well from the
overwhelming background from quantum chromodynamics (QCD).
The $A^0\to gg$ channel will greatly contribute to the decay mode to a
light-hadron dijet, which will be measurable at the ILC.

Since the $A^0$ boson is neutral and colorless, the $A^0\gamma\gamma$ and
$A^0gg$ couplings are loop induced.
As the $A^0$ boson has no tree-level coupling to the $W$ boson and its
coupling to sfermions flips their ``handedness'' (left or right), the
$A^0\gamma\gamma$ coupling is mediated at leading order (LO) by heavy quarks
and charged leptons and by light charginos \cite{Kalyniak:1985ct}.
The $A^0gg$ coupling is generated at LO by heavy-quark loops
\cite{Gunion:1986nh}.

Reliable theoretical predictions for the $A^0\gamma\gamma$ and $A^0gg$
couplings, including higher-order radiative corrections, are urgently required
to match the high precision to be reached by the LHC and ILC experiments
\cite{Battaglia:2000jb}.
Specifically, the properties of the $A^0$ boson, especially its CP-odd nature,
must be established, and the sensitivity to novel high-mass particles
circulating in the loops must be optimized.
The present state of the art is as follows.
The next-to-leading-order (NLO) QCD corrections,
of relative order ${\cal O}(\alpha_s)$ in the strong-coupling constant
$\alpha_s$, to the partial decay widths $\Gamma(A^0\to\gamma\gamma)$
\cite{Djouadi:1993ji,Spira:1995rr} and $\Gamma(A^0\to gg)$
\cite{Spira:1995rr}, and the production cross section $\sigma(gg\to A^0)$
\cite{Spira:1993bb,Spira:1995rr} are available for arbitrary values of quark
and $A^0$-boson masses as one-dimensional integrals, which were solved in
terms of harmonic polylogarithms for $\Gamma(A^0\to\gamma\gamma)$,
$\Gamma(A^0\to gg)$, and the virtual correction to $\sigma(gg\to A^0)$
\cite{Harlander:2005rq,Aglietti:2006tp}.
The latter was also obtained for general color factors of the gauge group
SU($N_c$) in the limit $m_t\to\infty$ using an effective Lagrangian
\cite{Ravindran:2004mb}.
The next-to-next-to-leading-order (NNLO) QCD corrections, of
${\cal O}(\alpha_s^2)$, to $\Gamma(A^0\to gg)$ \cite{Chetyrkin:1998mw}
and $\sigma(gg\to A^0)$ \cite{Harlander:2002vv} were found for $m_t\to\infty$
using an effective Lagrangian.
The ${\cal O}(\alpha_s)$ SUSY QCD correction, due to virtual squarks and
gluinos besides the heavy quarks, to $\sigma(gg\to A^0)$ was obtained from
an effective Lagrangian constructed by also integrating out the SUSY particles
\cite{Harlander:2005if}.
The two-loop master integrals appearing in the latter calculation if the
masses of the virtual scalar bosons and fermions are kept finite were
expressed in terms of harmonic polylogarithms \cite{Aglietti:2006tp}.

\begin{figure}[ht]
\begin{center}
\includegraphics[width=0.98\textwidth,viewport=120 474 477 724,clip]{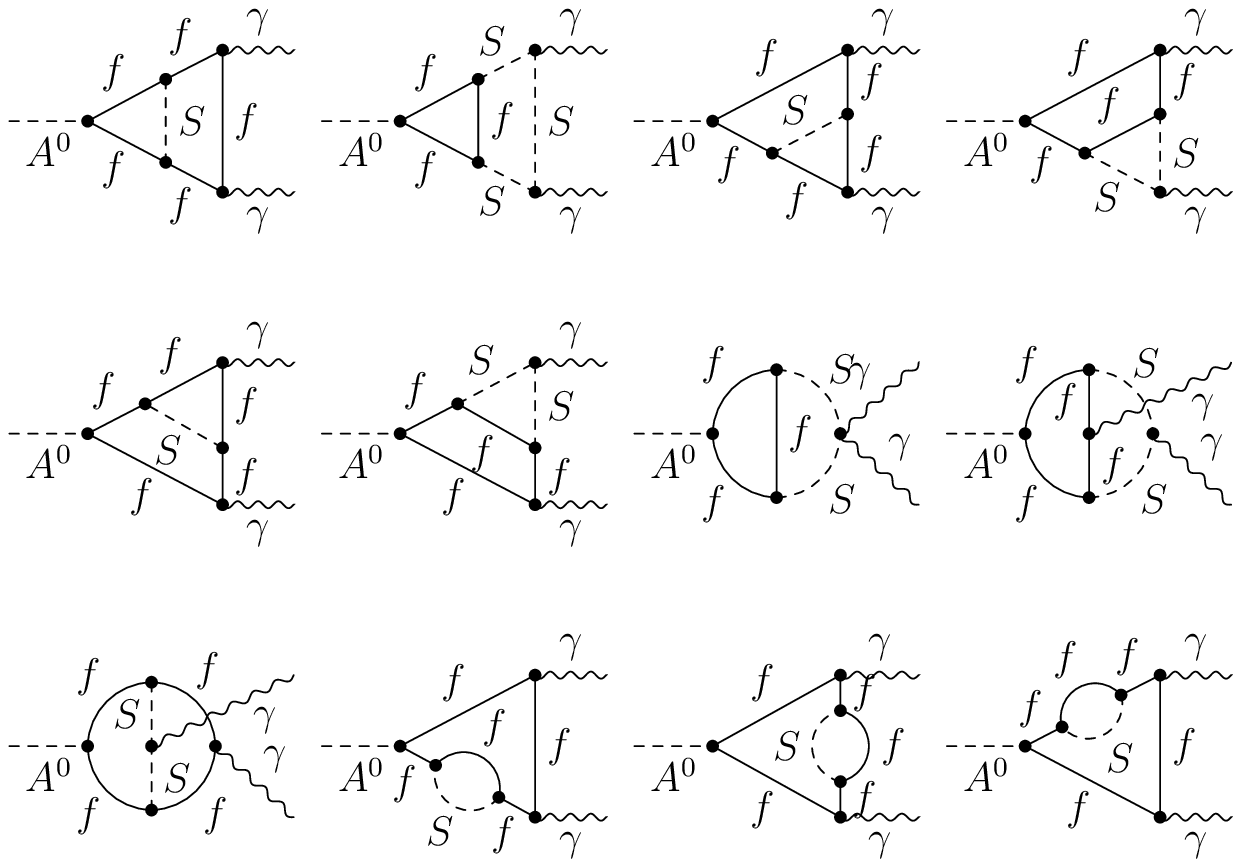}
\end{center}
\caption{\label{fig:one}%
Feynman diagrams contributing to $\Gamma(A^0\to\gamma\gamma)$ at
${\cal O}(x_t)$.
$S=w^\pm,z^0,h^0,H^0,A^0,H^\pm$ and $f=t,b$ denote generic scalar bosons and
fermions, respectively.
The couplings of the $z^0$, $h^0$, $H^0$, and $A^0$ bosons to the $b$ quark
are to be neglected and those to the photon vanish.}
\end{figure}

In this Letter, we take the next step and present the dominant electroweak
corrections to $\Gamma(A^0\to\gamma\gamma)$ and $\Gamma(A^0\to gg)$ at NLO.
Since these are purely virtual, arising from two-loop diagrams, they carry
over to $\sigma(\gamma\gamma\to A^0)$ and $\sigma(gg\to A^0)$, via
\begin{equation}
\sigma(\gamma\gamma/gg\to A^0)=\frac{8\pi^2}{N_{\gamma,g}^2M_{A^0}}
\Gamma(A^0\to\gamma\gamma/gg)\delta\left(\hat{s}-M_{A^0}^2\right),
\end{equation}
where $N_\gamma=1$ and $N_g=N_c^2-1=8$ are the color multiplicities of the
photon and the gluon, respectively, and $\hat{s}$ is the partonic c.m.\ energy
square.
For the time being, we focus our attention on the particularly interesting
region of parameter space with low to intermediate Higgs-boson masses,
$M_{h^0},M_{H^0},M_{A^0},M_{H^\pm}<m_t$, and low to moderate value of
$\tan\beta$, $\tan\beta\ll m_t/m_b$, and assume that the SUSY particles are so
heavy that they can be regarded as decoupled, yielding subdominant
contributions.
The dominant electroweak two-loop corrections are then of relative order
${\mathcal O}(x_t)$, where
$x_t=G_Fm_t^2/(8\pi^2\sqrt{2})\approx3.17\times10^{-3}$ with $G_F$ being
Fermi's constant.
In the case of $\Gamma(A^0\to\gamma\gamma)$, they arise from the class of
generic Feynman diagrams shown in Fig.~\ref{fig:one}, which, besides the $t$
and $b$ quarks, involve the charged and neutral Goldstone bosons, $w^\pm$ and
$z^0$, and the five Higgs bosons.
Here, it is understood that the $b$ quark only couples to the $w^\pm$ and
$H^\pm$ bosons because its couplings to the neutral scalar bosons are
suppressed unless $\tan\beta\alt m_t/m_b$ and, of course, that the $z^0$,
$h^0$, $H^0$, and $A^0$ bosons do not couple to the photon.
We explicitly checked that the $W^\pm$ and $Z^0$ bosons do not contribute at
${\mathcal O}(x_t)$.
The diagrams contributing to $\Gamma(A^0\to gg)$ at ${\mathcal O}(x_t)$ emerge
from Fig.~\ref{fig:one} by omitting those where a photon couples to a scalar
boson and by replacing the photons by gluons.

We now outline the course of our calculation and exhibit the structure of our
results.
Full details will be presented in a forthcoming communication \cite{long}.
Since we consider the SUSY partners to be decoupled, we may as well work in
the 2HDM without SUSY.
We may thus extract the ultraviolet (UV) divergences by means of dimensional
regularization, with $D=4-2\epsilon$ space-time dimensions and 't~Hooft mass
scale $\mu$.
For convenience, we work in 't~Hooft-Feynman gauge.
We take the Cabibbo-Kobayashi-Maskawa quark mixing matrix to be unity, which
is well justified because the third quark generation is, to good
approximation, decoupled from the first two.
We adopt Sirlin's formulation of the electroweak on-shell renormalization
scheme \cite{Sirlin:1980nh}, which uses $G_F$ and the physical particle masses
as basic parameters.
Various prescriptions for the renormalization of the auxiliary variable
$\tan\beta$, with specific virtues and flaws, may be found in the literature
(for a review, see Ref.~\cite{Freitas:2002um}).
For definiteness, we employ the Dabelstein-Chankowski-Pokorski-Rosiek (DCPR)
scheme \cite{Chankowski:1992er}, which maintains the relation
$\tan\beta=v_2/v_1$ in terms of the ``true'' vacua through the condition
$\delta v_1/v_1=\delta v_2/v_2$, and demands the residue condition
$\re\hat\Sigma_{A^0}^\prime(M_{A^0})=0$ and the vanishing of the $A^0$--$Z^0$
mixing on shell as $\re\hat\Sigma_{A^0Z^0}(M_{A^0})=0$, where
$\hat\Sigma_{A^0}(q^2)$ and $\hat\Sigma_{A^0Z^0}(q^2)$ are the renormalized
$A^0$-boson self-energy and $A^0$--$Z^0$ mixing amplitude, respectively.

\begin{figure}[ht]
\begin{center}
\includegraphics[width=0.98\textwidth,viewport=125 676 470 759,clip]{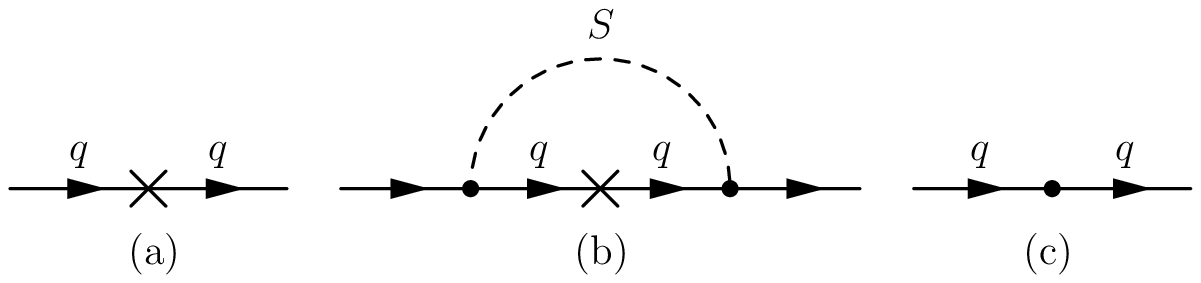}
\end{center}
\caption{\label{fig:two}%
Feynman diagrams contributing to $Z_5^p$ at ${\cal O}(x_t)$.
Crosses and dots indicate the insertions of the Fourier transform of $P(x)$
and its operator renormalization $Z_2Z^p$, respectively.}
\end{figure}

The evaluation of the diagrams in Fig.~\ref{fig:one} is aggravated by the
appearance of three $\gamma_5$ matrices inside closed fermion loops.
This leads us to adopt the 't~Hooft-Veltman-Breitenlohner-Maison (HVBM) scheme
\cite{tHooft:1972fi}, which allows for a consistent treatment of the Dirac
algebra within the framework of dimensional regularization.
Then, one has
\begin{equation}
\gamma_5=\frac{i}{4!}\varepsilon_{\mu\nu\rho\sigma}\,
\gamma^\mu\gamma^\nu\gamma^\rho\gamma^\sigma,
\end{equation}
where the totally antisymmetric Levi-Civita tensor is defined in $D$
dimensions as
\begin{equation}
\varepsilon_{\mu\nu\rho\sigma}=\left\{
\begin{tabular}{ll}
1&\mbox{if $(\mu,\nu,\rho,\sigma)$ even permutation of (0,1,2,3),}\\
-1&\mbox{if $(\mu,\nu,\rho,\sigma)$ odd permutation of (0,1,2,3),}\\
0&\mbox{otherwise.}
\end{tabular}\right.
\label{eq:epsilon}
\end{equation}
In fact, we explicitly verified that the na\"\i ve anticommuting definition of
the $\gamma_5$ matrix yields ambiguous results, which depend on the way of
executing the Dirac traces.
Furthermore, in the renormalization of the pseudoscalar current
\begin{equation}
P(x)=Z_2Z^pZ_5^p\overline{\psi}(x)\gamma_5\psi(x),
\end{equation}
one needs to introduce a finite renormalization constant $Z_5^p$, besides the
usual fermion wave-function and pseudoscalar-current UV renormalization
constants $Z_2$ and $Z^p$ of the modified minimal-subtraction
($\overline{\mathrm{MS}}$) scheme, to effectively restore the
anticommutativity of the $\gamma_5$ matrix \cite{Larin:1993tq}.
Within QCD, $Z_5^p$ is known through ${\cal O}(\alpha_s^3)$
\cite{Larin:1993tq}.
Here, we need $Z_5^p$ at ${\mathcal O}(x_t)$.
We thus need to consider the diagrams depicted in Fig.~\ref{fig:two} with
the external legs amputated, where a cross indicates the insertion of the
Fourier transform of $P(x)$ and a dot the operator renormalization.
Since the $A^0b\overline{b}$ coupling is suppressed in our case, dominant
contributions only arise from the neutral scalar bosons.
Using the mixed commutation and anticommutation relations properly
distinguishing between 4 and $(D-4)$ dimensions \cite{tHooft:1972fi}, we
decompose the string of gamma matrices appearing in the expression for
Fig.~\ref{fig:two}(b) into the term proportional to the LO result of
Fig.~\ref{fig:two}(a) that one would obtain with an anticommuting $\gamma_5$
matrix and an evanescent remainder, which lives in the unphysical
$(D-4)$-dimensional part of space-time and vanishes in the physical limit
$D\to4$.
Upon loop integration, the first term may produce an UV divergence, which
would be canceled by $Z_2Z^p$ in Fig.~\ref{fig:two}(c), while the evanescent
remainder may generate an unphysical finite contribution to be canceled by
$Z_5^p$.
By explicit evaluation, the latter is found to vanish at ${\mathcal O}(x_t)$,
owing to the cancellation of the individual contributions from the $z^0$,
$h^0$, $H^0$, and $A^0$ bosons, so that $Z_5^p=1$ for our application.

We first consider the $A^0\to\gamma\gamma$ decay.
By Lorentz covariance, its transition matrix element takes the form
\begin{equation}
{\mathcal T}=\varepsilon_{\mu\nu\rho\sigma}\varepsilon^{\mu*}(q_1)
\varepsilon^{\nu*}(q_2)q_1^\rho q_2^\sigma{\mathcal A},
\end{equation}
where $\varepsilon^\mu(q)$ is the polarization four-vector of a photon with
four-momentum $q^\mu$ and $\varepsilon_{\mu\nu\rho\sigma}$ is defined in
Eq.~(\ref{eq:epsilon}), so that
\begin{equation}
\Gamma(A^0\to\gamma\gamma)=\frac{M_{A^0}^3}{64\pi}|{\mathcal A}|^2.
\label{eq:gamma}
\end{equation}
The form factor ${\mathcal A}$ is evaluated perturbatively as
${\mathcal A}={\mathcal A}_0+{\mathcal A}_{\alpha_s}+{\mathcal A}_{x_t}
+\cdots$.
In $D$ dimensions, the LO result reads
\begin{equation}
{\mathcal A}_0=-CQ_t^2
\left(\frac{4\pi\mu^2}{m_t^2}\mathrm{e}^{-\gamma_E}\right)^\epsilon
\left[\frac{1}{\tau}\arcsin^2\sqrt{\tau}+{\mathcal O}(\epsilon)\right],
\label{eq:born}
\end{equation}
where $C=(2^{1/4}/\pi)G_F^{1/2}\alpha N_c\cot\beta$ with $\alpha$ being
Sommerfeld's fine-structure constant, $Q_t=2/3$ is the fractional electric
charge of the top quark, $\gamma_E$ is the Euler-Mascheroni constant, and
$\tau=M_{A^0}^2/(2m_t)^2$.
For $\tau\ll 1$, the function within the square brackets of
Eq.~(\ref{eq:born}) has the expansion $1+{\mathcal O}(\tau)$.
The ${\mathcal O}(x_t)$ result
${\mathcal A}_{x_t}={\mathcal A}_{x_t}^\mathrm{CT}+{\mathcal A}_{x_t}^0$ is
composed of a counterterm ${\mathcal A}_{x_t}^\mathrm{CT}$ and the
contribution ${\mathcal A}_{x_t}^0$ from the proper vertex diagrams in
Fig.~\ref{fig:one}.
We have
\begin{equation}
{\mathcal A}_{x_t}^\mathrm{CT}=-CQ_t^2Z_5^p\left(\frac{\delta v}{v}
-\frac{\Delta r}{2}-2\epsilon\frac{\delta m_t}{m_t}\right),
\label{eq:ct}
\end{equation}
where $\delta v/v$ is the common DCPR counterterm for the two Higgs doublets
given in Eq.~(3.11) of Ref.~\cite{Dabelstein:1995js}, $\Delta r$
\cite{Sirlin:1980nh} contains those radiative corrections to the muon lifetime
which the SM introduces on top of those derived in the QED-improved Fermi
model, and $\delta m_t/m_t$ may be found, {\it e.g.}, in Eq.~(74) of
Ref.~\cite{Butenschoen:2007hz}.
In terms of (transverse) self-energies, we have
\begin{eqnarray}
\frac{\delta v}{v}-\frac{\Delta r}{2}&=&\frac{1}{2}
\left[-\frac{\Sigma_{W^\pm,T}(0)}{M_W^2}-\Sigma_{A^0}^\prime(M_{A^0}^2)
+(\tan\beta-\cot\beta)
\frac{\Sigma_{A^0Z^0}(M_{A^0}^2)}{M_Z}\right]
\nonumber\\
&=&\frac{N_c}{2}x_t.
\end{eqnarray}
We evaluate ${\mathcal A}_{x_t}^0$ by applying the asymptotic-expansion
technique with the help of the programs \texttt{QGRAF} \cite{Nogueira:1991ex},
\texttt{q2e}, \texttt{exp} \cite{Harlander:1997zb}, and \texttt{MATAD}
\cite{Steinhauser:2000ry}.
Our final result reads
\begin{equation}
{\mathcal A}_{x_t}=Cx_t\frac{2}{9}\left(\frac{7}{\sin^2\beta}-N_c\right).
\label{eq:ew}
\end{equation}
Comparison with Eq.~(\ref{eq:born}) shows that, for $\tau\ll1$,
$\Gamma(A^0\to\gamma\gamma)$ receives the electroweak correction factor
$[1-x_t(4+7/\tan^2\beta)]$.

We now turn to $\Gamma(A^0\to gg)$.
We then need to include the color factor $N_g/4=2$ in Eq.~(\ref{eq:gamma}) and
substitute $C\to\tilde{C}=(2^{1/4}/\pi)G_F^{1/2}\alpha_s\cot\beta$ and
$Q_t\to1$ in Eqs.~(\ref{eq:born}) and (\ref{eq:ct}).
Implementing the appropriate substitutions in the relevant subset of diagrams
in Fig.~\ref{fig:one} and combining the outcome with the counterpart of
Eq.~(\ref{eq:ct}), the counterpart of Eq.~(\ref{eq:ew}) is found to be
\begin{equation}
\tilde{{\mathcal A}}_{x_t}=\tilde{C}x_t\left(\frac{5}{\sin^2\beta}
-\frac{N_c}{2}\right),
\end{equation}
so that $\Gamma(A^0\to gg)$ receives the electroweak correction factor
$[1-x_t(7+10/\tan^2\beta)]$.

\begin{figure}[ht]
\begin{center}
\begin{tabular}{cc}
\includegraphics[width=0.48\textwidth,viewport=19 20 638 435,clip]{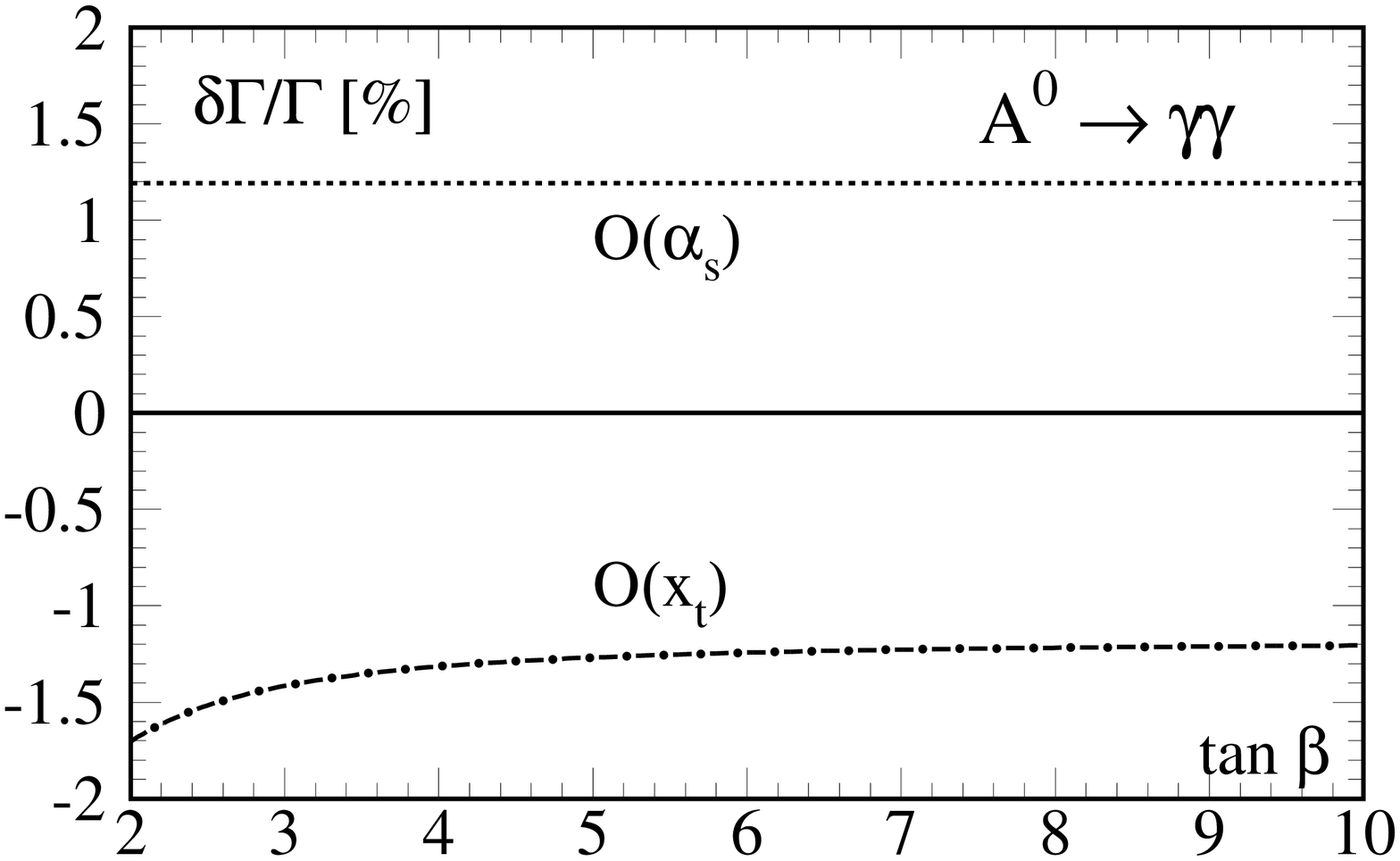}
&
\includegraphics[width=0.48\textwidth,viewport=19 20 638 435,clip]{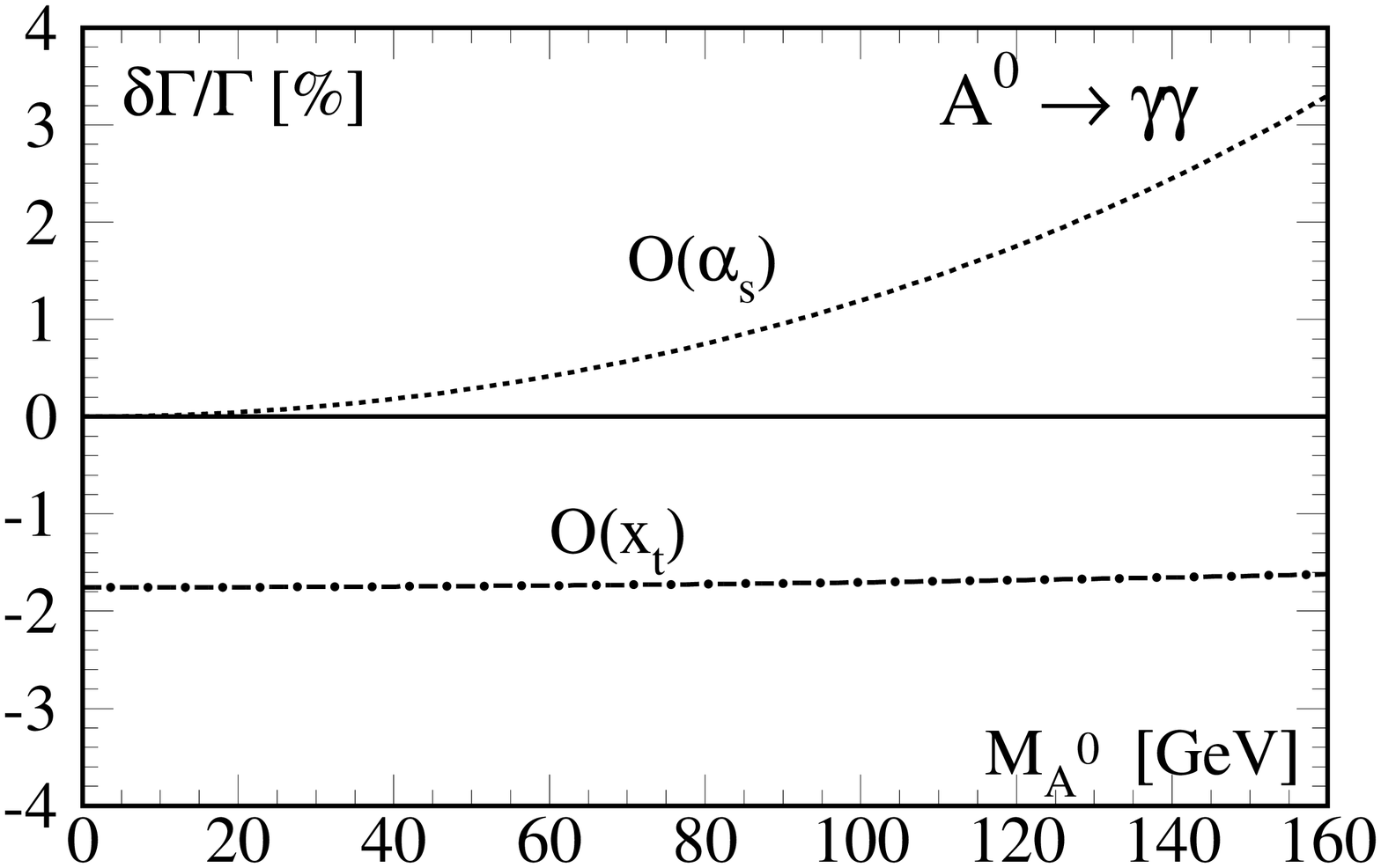}
\end{tabular}
\end{center}
\caption{\label{fig:three}%
${\mathcal O}(x_t)$ and ${\mathcal O}(\alpha_s)$ corrections to
$\Gamma(A^0\to\gamma\gamma)$ (a) for $M_{A^0}=100$~GeV as functions of
$\tan\beta$ and (b) for $\tan\beta=2$ as functions of $M_{A^0}$.}
\end{figure}
As a check for our computational setup, we also recalculated the 
${\mathcal O}(\alpha_s)$ corrections to $Z_5^p$ and, as an expansion in
$\tau$ through ${\mathcal O}(\tau^4)$, to $\Gamma(A^0\to\gamma\gamma)$ to find
agreement with Refs.~\cite{Larin:1993tq} and
\cite{Djouadi:1993ji,Spira:1995rr,Harlander:2005rq,Aglietti:2006tp},
respectively.
Notice that the ${\mathcal O}(\tau^0)$ term vanishes, so that the 
${\mathcal O}(\alpha_s)$ correction is suppressed for small values of
$M_{A^0}$.
In fact, as a consequence of the Adler-Bardeen theorem \cite{Adler:1969er}, 
the large-$m_t$ effective Lagrangian of the $A^0\gamma\gamma$ interaction does
not receive QCD corrections at any order \cite{Kniehl:1995tn}.
The ${\mathcal O}(x_t)$ and ${\mathcal O}(\alpha_s)$ corrections to
$\Gamma(A^0\to\gamma\gamma)$ are compared in Fig.~\ref{fig:three}.
We observe from Fig.~\ref{fig:three}(a) that the ${\mathcal O}(x_t)$ 
correction amounts to $-1.7\%$ at $\tan\beta=2$ and rapidly reaches its
asymptotic value of $-1.2\%$ as $\tan\beta$ increases, whereas the
${\mathcal O}(\alpha_s)$ correction is positive and independent of
$\tan\beta$.
The $M_{A^0}$ dependence of the ${\mathcal O}(x_t)$ correction shown in
Fig.~\ref{fig:three}(b) is induced by ${\mathcal A}_0$ in Eq.~(\ref{eq:born})
to which ${\mathcal A}_{x_t}$ is normalized.
Since it is rather feeble, we may expect the unknown ${\mathcal O}(\tau^n)$
($n=1,2,3,\ldots$) terms in Eq.~(\ref{eq:ew}) to be of moderate size, too.
The smallness and approximately quadratic $M_{A^0}$ dependence of the
${\mathcal O}(\alpha_s)$ correction is due to the absence of the leading
${\mathcal O}(\tau^0)$ term discussed above.
We conclude that the ${\mathcal O}(x_t)$ reduction more than compensates the
${\mathcal O}(\alpha_s)$ enhancement for $M_{A^0}\alt120$~GeV.

The ${\mathcal O}(x_t)$ correction to $\Gamma(A^0\to gg)$ ranges from $-2.8\%$
at $\tan\beta=2$ to the asymptotic value $-2.1\%$ and partly screens the
sizeable ${\mathcal O}(\alpha_s)$ and ${\mathcal O}(\alpha_s^2)$ corrections
of about $68\%$ and $23\%$, respectively, which do have non-vanishing
${\mathcal O}(\tau^0)$ terms \cite{Spira:1995rr,Chetyrkin:1998mw}.

In conclusion, we analytically calculated the dominant electroweak two-loop
corrections, of order ${\cal O}(x_t)$, to $\Gamma(A^0\to\gamma\gamma)$,
$\Gamma(A^0\to gg)$, $\sigma(\gamma\gamma\to A^0)$, and $\sigma(gg\to A^0)$
within the 2HDM with low- to intermediate-mass Higgs bosons for small to
moderate value of $\tan\beta$ using asymptotic expansion in
$M_{A^0}^2/(2m_t)^2$.
To consistently overcome the non-trivial $\gamma_5$ problem of dimensional
regularization, we adopted the HVBM scheme and included the finite
renormalization constant $Z_5^p$ of the pseudoscalar current to effectively
restore the anticommutativity of the $\gamma_5$ matrix.
The ${\cal O}(x_t)$ term of $Z_5^p$ was found to vanish.
On the phenomenological side, the ${\cal O}(x_t)$ correction to
$\Gamma(A^0\to\gamma\gamma)$ and $\sigma(\gamma\gamma\to A^0)$ is of major
importance, since it more than compensates the ${\mathcal O}(\alpha_s)$
enhancement for $M_{A^0}\alt120$~GeV.
As for $\Gamma(A^0\to gg)$ and $\sigma(gg\to A^0)$, the ${\cal O}(x_t)$
correction appreciably screens the sizeable QCD enhancement, by up to $-3\%$.

We thank W. Hollik for a useful communication concerning the renormalization
of $\tan\beta$, P. Kant for providing us with the series expansion of
Eq.~(2.20) in Ref.~\cite{Harlander:2005rq}, and M. Gorbahn, M. Spira, and
M. Steinhauser for beneficial discussions.
This work was supported in part by BMBF Grant No.\ 05~HT6GUA and DFG Grant
No.\ GRK~602.

\end{document}